# Effect of Electron Irradiation on the Transport and Field Emission Properties of Few-Layer $MoS_2$ Field Effect Transistors

Filippo Giubileo, Laura Iemmo, Maurizio Passacantando, Francesca Urban, Giuseppe Luongo, Lingfeng Sun, Giampiero Amato, Emanuele Enrico, Antonio Di Bartolomeo*

*CNR-SPIN Salerno, via Giovanni Paolo II n. 132, Fisciano 84084, Italy*
*Physics Department, University of Salerno, and CNR-SPIN, via Giovanni Paolo II n. 132, Fisciano 84084, Italy*
*Department of Physical and Chemical Science, University of L'Aquila, and CNR-SPIN L'Aquila, via Vetoio, Coppito 67100, L'Aquila, Italy*
*Department of Energy Science, Sungkyunkwan University, Suwon 16419, Korea*
*Istituto Nazionale di Ricerca Metrologica, INRIM - Strada delle Cacce, Torino 10135, Italy*

**Abstract**

Electrical characterization of few-layer $MoS_2$ based field effect transistors with Ti/Au electrodes is performed in the vacuum chamber of a scanning electron microscope in order to study the effects of electron beam irradiation on the transport properties of the device. A negative threshold voltage shift and a carrier mobility enhancement is observed and explained in terms of positive charges trapped in the $SiO_2$ gate oxide, during the irradiation. The transistor channel current is increased up to three order of magnitudes after the exposure to an irradiation dose of $100 e^-/nm^2$. Finally, a complete field emission characterization of the $MoS_2$ flake, achieving emission stability for several hours and a minimum turn-on field of $\approx 20$ V/μm with a field enhancement factor of about 500 at anode-cathode distance of ~1.5 μm, demonstrates the suitability of few-layer $MoS_2$ as two-dimensional emitting surface for cold-cathode applications.

1. **Introduction**

Molybdenum disulfide ($MoS_2$) is a two-dimensional (2D) layered material, one of the transition-metal dichalcogenides (TMDs), with layers that are weakly held together by van der Waals forces. Energy band gap in $MoS_2$ varies from 1.2 eV (indirect) in the bulk to 1.8-1.9 eV (direct) in monolayer.[1-3] Despite a low field-effect mobility (limited to few hundreds $cm^2V^{-1}s^{-1}$ on suspended samples),[4] $MoS_2$-based devices have attracted growing interest for several applications, such as field-effect transistors (FETs),[2,5-9] sensors,[10-12] spintronic devices,[13] field emission cathodes,[14-15] synaptic computation for neuroscience,[16] etc.

The development of $MoS_2$-based nanoelectronics needs to overcome the difficulties arising from point defects as well as structural damages and dislocations, often generated during the fabrication processes. For instance, structural defects behave as charge traps modifying the electronic properties of devices.[17] Nowadays, the device fabrication and characterization activities are intimately related to the application of scanning electron microscopy (SEM), electron beam lithography (EBL), transmission electron microscopy (TEM), and focus ion beam (FIB) processing, which can have non-negligible effects due to the exposure to electrons or ions bombardment. Indeed, it has been demonstrated that the irradiation by energetic particles (electrons and/or ions) can provoke relevant modifications of the electronic properties of 2D materials by introducing damage and/or defects.[18-21] On the other hand, electrical and optical device properties may be modified by intentionally creating defects by means of electron-beam [17] or ion irradiation [22] as well as by plasma treatments.[23]

Among the studies available about the effect of irradiation on 2D materials, Komsa *et al.* used first-principles atomistic simulations to study the response of TMDs layers to electron irradiation.[24] They calculated displacement threshold energies for atoms in several compounds, including $MoS_2$, and gave the corresponding electron energies necessary to produce defects. They also performed high-resolution TEM experiments on $MoS_2$, reporting that e-beam energy of about 90 keV is effective to produce sulphur vacancies by knock-on mechanism. Choi *et al.* studied the effects of 30 keV electron-beam irradiation on monolayer $MoS_2$ FETs, reporting that irradiation-induced defects act as trap sites which reduce the carrier mobility and concentration while shifting the threshold voltage.[25] Zhou *et al.* performed a systematic study of point defects in $MoS_2$ using both SEM imaging and first-principles calculations demonstrating that vacancies are created by e-beam irradiation at low energies (~30 keV).[26] Durand *et al.* studied the effects of e-beam irradiation on the transport properties of CVD-grown $MoS_2$ in FET configuration reporting an increase of the carrier density and a reduction of the mobility explained as the consequence of both intrinsic defects in $MoS_2$ and Coulomb potential of irradiation induced charges at the $MoS_2/SiO_2$ interface.[27] Recently, it has also been demonstrated that e-beam irradiation on $MoS_2$ based FET can produce a negative threshold voltage shift followed by a positive shift by increasing the aging time.[28]

Higher energetic irradiation on $MoS_2$ has been also tested. Ochedowski *et al.* used 1.14 GeV $U^{28+}$ ion beam showing that conductivity of FETs is deteriorated by irradiation up to a complete destruction of the device for a fluence as high as $4\times10^{11}$ ions/$cm^2$.[29] Effects of irradiation by 10 MeV proton beams has been reported by Kim *et al.* for fluence up to $10^{14}$ $cm^{-2}$.[30] They demonstrated that electrical properties were unaffected for fluence up to $10^{12}$ $cm^{-2}$, while higher values caused a reduction of current level and of the conductance as well as a shift of the threshold voltage toward the positive gate voltage. The main mechanism was identified in the formation of irradiation-induced traps, such as positive oxide-charge traps in the $SiO_2$ layer and interface trap states. Moreover, recovery of such modifications has been experimentally proved over a time scale of few days.

In this paper, we perform a systematic electrical characterization of CVD synthesized few-layer $MoS_2$ based FETs, inside a scanning electron microscope, to study the effects of low energy (up to 10 keV) e-beam irradiation. We report an increase of the carrier mobility and a negative shift of the threshold voltage for successive low energy irradiations that is explained in terms of positive charge trapped in the $SiO_2$ gate dielectric.

Moreover, taking advantage of the measurement setup with nano-controlled metallic tips inside the SEM chamber, we also perform a complete characterization of the field emission properties of the few-layer MoS$_2$. Indeed, due to the intrinsically sharp edges and high aspect ratio of MoS$_2$, this layered material is a natural candidate to realize high performance field emission cathodes. We demonstrate a turn-on field of ≈ 20 V/μm and a field enhancement factor of about 500, at anode-cathode distance of about 1.5 μm. Finally, we show that MoS$_2$ allows high current emission with high time stability, with fluctuations of the order of 5%.

## 2. Methods

MoS$_2$ flakes were synthesized on SiO$_2$(300nm)/p-Si substrate by chemical vapor deposition at high temperature, using sulfur (S) and molybdenum trioxide (MoO$_3$) powders as solid precursors. The growth process was performed in Ar atmosphere (50 sccm gas flow); temperature was raised with fixed rate of 20°C/min from room temperature to 850 °C, and then kept for 10 min for the material growth. Finally the substrate was left to cool down naturally. Micro-Raman spectroscopy measurements with a 532 nm laser source were performed on the selected flakes in order to precisely identify the number of layers. The laser power was kept below 0.5 mW in order to avoid heating and/or modifications of the flakes. **Figure 1**(a) shows a typical Raman spectrum measured on one of the flakes, showing the $E_{2g}^1$ and $A_{1g}$ peak, which is due to the in-plane and out-of-plane vibrations of Mo and S atoms, respectively. The frequency separation of the two peaks of about 23-24 cm$^{-1}$ indicates a few-layer flake.[31] A scanning electron microscopy (SEM) image of the flake is reported as inset of Figure 1(a). Two metal contacts, deposited by standard electron-beam lithography (EBL) and lift-off process, are visible: a larger one of Ti(20 nm)/Au(130 nm) and a shorter one, cross shaped, of Au (130 nm).

In Figure 1(b), we show a schematic layout of the device and of the experimental setup. Electrical measurements were performed inside a Zeiss LEO 1430 SEM chamber in high vacuum (pressure lower than $10^{-6}$ Torr) and at room temperature. Two tungsten tips, mounted on a nanoprobes system with two piezoelectric-driven arms installed inside the SEM chamber, were electrically connected to a semiconductor parameter analyser (Keithley 4200-SCS) working as source-measurement unit (SMU), to apply bias (up to ±120 V) and to measure the current with sensibility of about $10^{-14}$ A. The circuit configuration for the field effect transistor was obtained by using the silicon substrate as common back gate and the two metal leads as the drain and the source.

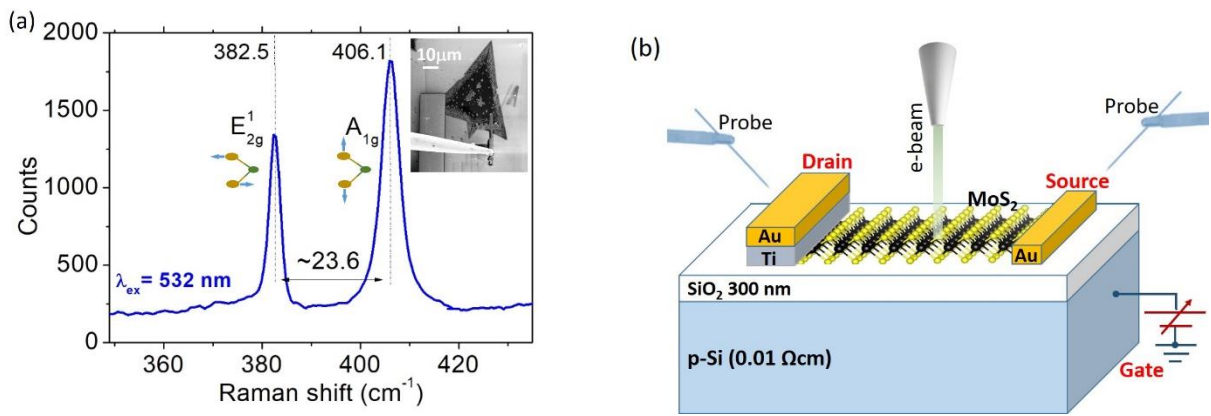

**Figure 1.** (a) Raman spectrum of few-layer $MoS_2$ flake measured for the flake imaged in the inset by using a laser source with 532 nm excitation wavelength. The atomic displacement of the two representative Raman-active modes ($E^1_{2g}$ and $A_{1g}$) are also shown. (b) Schematic layout of the device and of the measurement setup.

## 3. Results and discussion

### 3.1. Transistor characterization.

**Figure 2**(a) shows the output characteristics $I_{ds}-V_{ds}$ (plotted in logarithmic scale), measured in the two-probe configuration for the $MoS_2$ back-gated FET of Figure 1(b), performing a $V_{ds}$ voltage sweep from -3V to +3V and repeating the measurements for different gate voltages in the range -45V<$V_{gs}$<+45V. The characteristics indicate ohmic behaviour at low voltage, while for higher $V_{ds}$ voltages and high negative $V_{gs}$ bias a small asymmetry is observed (see inset). Similar non-linear

characteristics have been already discussed in terms of asymmetric Scottky contacts,[32,33] and are probably caused here by the use of Ti/Au and Au as metal leads.

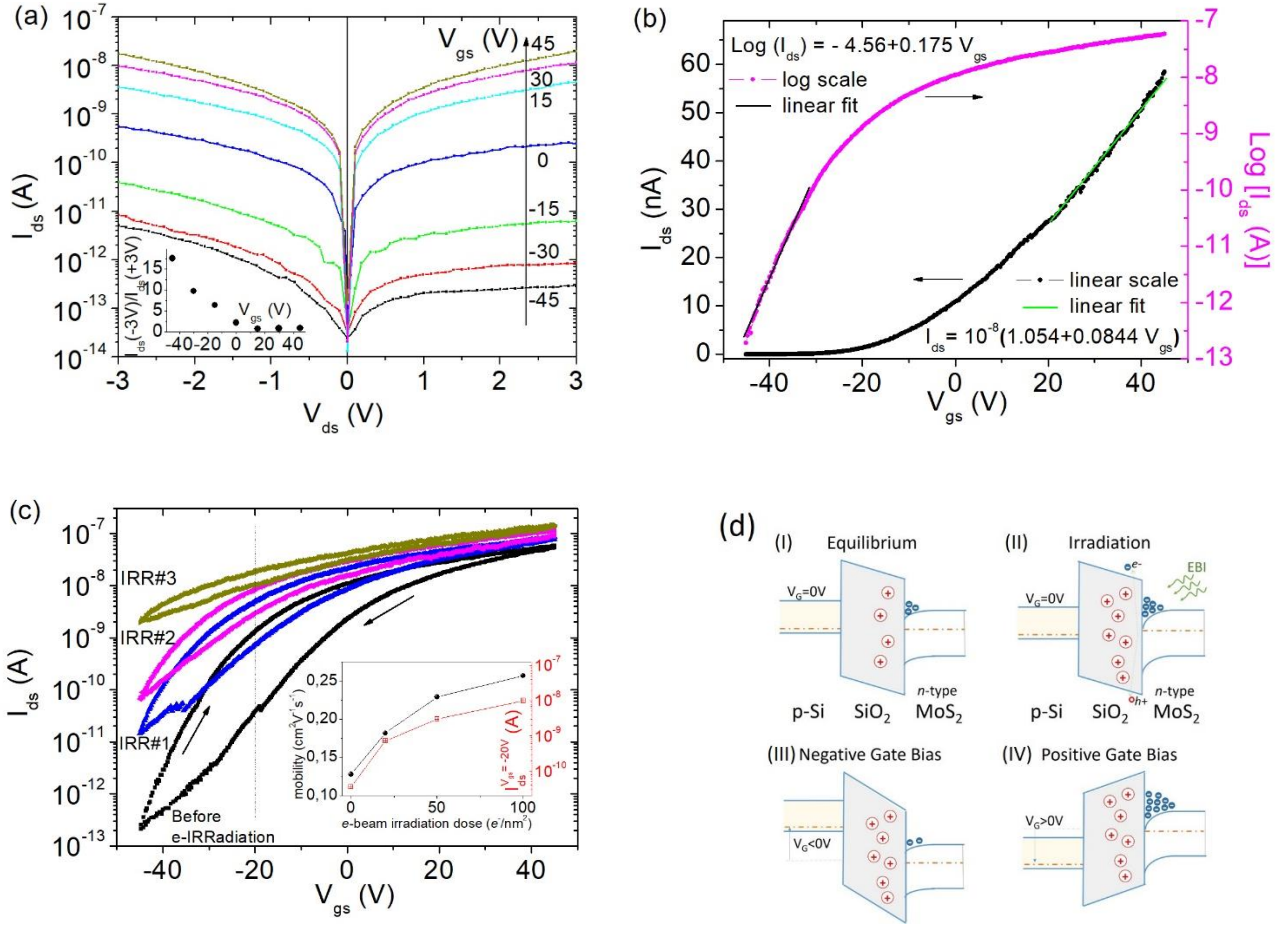

**Figure 2.** (a) $I_{ds} - V_{ds}$ output characteristics of the MoS$_2$ back-gated FET at different gate voltages. Inset: $I_{ds}$(-3V)/$I_{ds}$(+3V) ratio to quantify the slightly asymmetry arising at high negative gate voltages; (b) Transfer characteristic $I_{ds} - V_{gs}$ measured at $V_{ds}$ = 1.6 V. Curve is reported in both linear (left axis) and logarithmic scale (right axis). Results of linear fits are also shown; (c) Transfer characteristics (complete sweep loop +45V -> -45V -> +45V) $I_{ds} - V_{gs}$ measured at $V_{ds}$ = 1.6 V before and after electron beam irradiation. The irradiation dose is increased for each successive cycle. Inset: dependence of channel current and of carrier mobility on the irradiation dose; (d) Schematic band diagram for the n-type MoS$_2$/SiO$_2$/p-Si FET. (I) unbiased initial state; (II) unbiased state after irradiation which causes electron-hole pairs formation in SiO$_2$ and favours the formation of additional positive charged traps; (III) band alignment for $V_{gs}$<0V with carrier depleted channel; (IV) band alignment for $V_{gs}$>0V with carrier accumulation.

In Figure 2(b) we report the transfer characteristic $I_{ds} - V_{gs}$ at the drain bias $V_{ds}$ = 1.6 V, in linear (black curve, left scale) and logarithmic scale (blue curve, right scale) evidencing a threshold voltage $V_{th}$ = -21.5 V (here defined as the gate voltage to obtain a channel current $I_{ds}$ = 1 nA), i.e. that we are dealing with a n-type MoS2 based FET. The n-doping of few-layer MoS2 is often ascribed to the chemisorption of oxygen molecules on surface defects of MoS2, which locally lowers the conduction band edge, so promoting the n-doping of MoS2, but with no significant effect on the mobility and on the On/Off ratio of the transistor.[6,34,35]

The carrier mobility for the device under investigation is estimated using the equation

$$\mu = \frac{L}{W} \frac{1}{C_{SiO_2}} \frac{1}{V_{ds}} \frac{dI_{ds}}{dV_{gs}}$$

where $L$ and $W$ are the geometrical parameters of the transistor (length and width of the channel, respectively), and $C_{SiO_2}$ is the capacitance per unit area of the SiO2 gate dielectric ($C_{SiO_2} = (\epsilon_0 \cdot \epsilon_{SiO_2})/t_{SiO_2} = 1.15 \cdot 10^{-4}$ F/m², $\epsilon_0$ is the vacuum permittivity, $\epsilon_{SiO_2}$=3.9 and $t_{SiO_2}$= 300 nm are the relative permittivity and the thickness of SiO2, respectively). Using the slope of the $I_{ds} - V_{gs}$ curve in the linear region, we obtain the intrinsic field-effect mobility, $\mu = 0.12$ cm²V⁻¹s⁻¹, a value within the typically reported range 0.01-100 cm²V⁻¹s⁻¹ for CVD-grown MoS2 based FETs on thermally grown SiO2.[36,37] The observed low conductivity (on-state conductivity $\sigma_{ON} = (I_{ds}/V_{ds}) \cdot (L/W) \approx 2nS$) is originated by the relative high contact resistance due to the Schottky barriers at the contacts.[33,38] The low mobility is representative of a high density of scatterers, such as charged impurities due to the fabrication process and/or exposure to air, or intrinsic defects due to high surface-to-volume ratio of MoS2.

From the logarithmic plot of $I_{ds} - V_{gs}$ we can also evaluate the On/Off ratio (greater than $10^5$) and the sub-threshold swing $SS$, i.e. the gate voltage change required to increase the current in the transistor channel by one decade. In conventional FETs, $SS$ depends on the MOS capacitances as $SS = \frac{dV_{gs}}{d(\log(I_{ds}))} \approx \ln(10) \frac{kT}{q} \left(1 + \frac{C_T + C_{DL}}{C_{SiO_2}}\right)$, where $k$ is the Boltzmann constant, $T$ is the temperature, $q$ is the electron charge, $C_{DL}$ is the depletion layer capacitance, and $C_T$ is the capacitance associated with the interfacial charge traps. From the experimental curve $Log(I_{ds})$ $vs$ $V_{gs}$ in Figure 2(b) we obtain $SS = 5.7$V/decade. The relatively high value of $SS$ (with respect the minimum value $SS_{Min} = $

$\ln(10)\frac{kT}{q} \approx 60$ mV/decade for the ideal metal-oxide-semiconductor field-effect transistor) gives indication that $C_{SiO_2}$ is ten to hundred times smaller with respect the other capacitances. On the other hand, $C_{DL}$ is not expected to be a significant fraction of the total capacitance, due to the small sample thickness with respect the oxide thickness $t_{SiO_2}$. Consequently, we can give an estimation of the density of trap states $D_{Trap} \approx 7 \cdot 10^{12}$ cm$^{-2}$eV$^{-1}$ considering that $C_T = q^2 D_{Trap}$. This is a reasonable value in agreement with existing data.[39]

In the following we discuss the effect of electron beam irradiation (EBI) on the transfer characteristic of the MoS$_2$ transistor. All measurements were performed in-situ in the SEM chamber soon after the exposure to e-beam in order to avoid competing effects due to the air exposition of the device. The irradiation is performed at the electron beam energy of 10 keV, and at the fixed beam current of 0.2 nA. In Figure 2(c) we show complete (forward and backward) voltage sweeps in the range -45V< $V_{gs}$ <+45V. The first measurement performed before the irradiation is compared to the measurements obtained after irradiation for three different levels of electron irradiation dose in the range up to 100 $e^-$/nm$^2$. We observe that the threshold voltage is shifted towards more negative voltages after each irradiation (larger negative shift correspond to higher dose of irradiation). To give a quantitative estimation we consider the current flowing in the FET channel $I_{ds}$ (measured in the backward sweep at $V_{gs}$=-20V), that increases for each successive EBI cycle for increasing irradiation dose. In Figure 2(d) we summarize the resulting evolution of the channel current and of the carrier mobility due to EBI. We clearly see that both, $I_{ds}$ and $\mu$, monotonously raise for increasing dose. The maximum dose of 100 $e^-$/nm$^2$ almost causes a channel current increase by about three order of magnitude and a doubled mobility ($\mu \approx 0.25$ cm$^2$V$^{-1}$s$^{-1}$). These observations can be ascribed to the low energy of irradiation. Indeed, it has been reported that higher e-beam energy (30keV and a dose of $5 \cdot 10^2$ $e^-$/nm$^2$) is necessary to intentionally create defect sites (as mono-sulfur vacancies) causing a reduction of the channel current as well as of the carrier mobility in MoS$_2$ FET.[17,25,40] Defect sites may also be passivated (by chemisorption of C$_{12}$ molecules) obtaining an improvement of the device performance towards pre-irradiation values.[22,25] Moreover, a deep theoretical and experimental study of the effects of electron irradiation on few-layer MoS$_2$ flakes has demonstrated that only beam energies above 20 keV can systematically cause a decrease of channel conductivity in the transistor, while lower energies always result in an increase of conductivity,[40] confirming that electron irradiation can also be suitable to improve physical properties of MoS$_2$ based FETs.

The observed negative shift of the threshold voltage (and increased channel current) is explained by the pile up of positive charge in the SiO$_2$ traps that cause an enhancement of the gate electric field and an increase of carrier concentration.[41] Indeed, the EBI produces electron-hole pairs in the SiO$_2$ gate oxide, and due to the higher electron mobility, negative charges can rapidly escape, while holes are trapped.[41] Moreover, the induced EBI charges can also produce the formation of interfacial positive trapped charge that would contribute to the observed effect. A schematic band diagram of the underlying physical mechanism to explain the observed behaviour of n-type MoS$_2$ FET on p-Si/SiO$_2$ under EBI is shown in Figure 2(d). Configuration (I) represents the equilibrium state: electrons flow from MoS$_2$ to the interface due to the higher Fermi energy level of the MoS$_2$ till Fermi levels are aligned. The main effect of EBI is the formation of electro-holes pairs in the SiO$_2$ oxide (within first 100 nm layer), and the formation of further positive oxide trapped charges (II). Actually, trapped positive oxide charges may be already present in the oxide and (contribute to the n-doping of MoS$_2$), due to fabrication process as well as due to initial SEM imaging of the device. According to band diagrams (III) and (IV) depicted in Figure 2(d), when applying positive gate bias ($V_{gs}$>0V), electrons are attracted to the interface between MoS$_2$ and SiO$_2$ to form an accumulation layer. Vice versa, for negative gate bias ($V_{gs}$<0V), electrons are depleted from the channel.

### 3.2. Field emission characterization.

The circuit configuration for field emission characterization has been easily obtained by retracting one probe and finely adjusting its distance $d$ from the MoS$_2$ surface, the second probe contacting one metallic pad (inset of **Figure 3**(a)). The cathode($MoS_2$)-anode(suspended W-tip) separation $d$ can be precisely tuned with step resolution down to 5 nm. The use of a tip-shaped anode is an effective technique to perform FE characterization of reduced emitting areas (below 1 $\mu m^2$) with respect the standard parallel-plate setup that typically probes larger areas of several $mm^2$.[42-45]

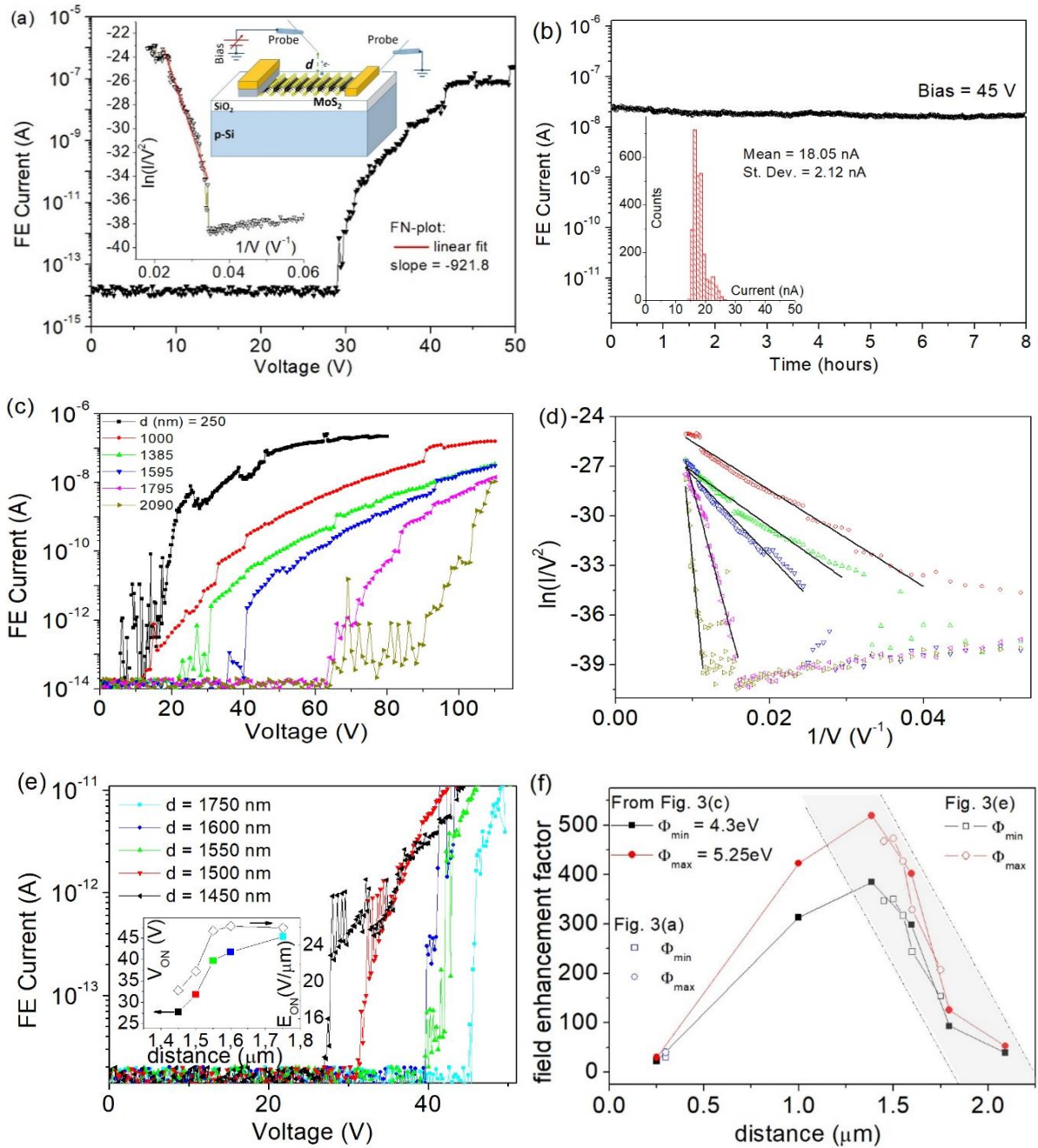

**Figure 3.** Field emission characterization of $MoS_2$ flake. (a) I-V curve measured at cathode-anode separation d =300nm. Left Inset: FN-plot of the experimental data. Red line is the linear fit. Upper inset: Schematic of the setup for FE measurements. (b) Field emission current stability measured at fixed voltage of 45 V. The inset show the histogram of the measured values. Mean and standard deviation are also reported. (c) FE characteristics measured in a second location on the $MoS_2$ surface

for several d values and corresponding (d) FN-plots (the plot for $d$ =250nm is not reported for clarity, being too noisy). (e) FE characteristics measured in a third location on the MoS$_2$ surface for a reduced range of the cathode-anode separation $d$ to precisely extract the turn-on field (voltage) vs $d$ (in the inset). (f) Dependence of the field enhancement factor on the distance $d$ as extracted from all the I-V curves reported and by considering two possible values of the work function ($\Phi_{min}$=4.3eV and $\Phi_{max}$=5.25eV).

We remark that the FE curves typically show large instabilities (fluctuations and drops) due to the presence and desorption of adsorbates (on the emitting surface), which act as nanoprotrusions with higher field enhancement factor and can be evaporated by Joule heating for the high FE currents.[46,47] Consequently, as standard procedure, we always perform an electric conditioning by repeating several successive voltage sweeps (not reported here) to stabilize the emitting surface. All reported data in the following have been measured after proper electric conditioning.

In Figure 3(a) we show a typical I-V curve measured at a cathode-anode separation $d = 300$ nm, where a turn-on field $E_{ON} = V_{ON}/d = 90$V/µm is necessary to start the current emission that rapidly increase for seven order of magnitudes from the setup floor noise of about 10$^{-14}$ A in about 20V bias range starting from the $V_{ON} \approx 30$ V. Moreover, when using a tip-shaped anode, a more precise estimation of the turn-on field $E_{ON}^*$ is obtained by considering a correction factor $k_{tip} \approx 1.5$,[42] resulting in a lower turn-on field value $E_{ON}^* = E_{ON}/k_{tip} \approx 60$V/µm.

Experimental data are then analysed in the framework of Fowler-Nordheim (FN) theory to verify the FE nature of the measured current, indeed it should follow the relation:[48]

$$I = a \frac{E_L^2}{\Phi} S \cdot exp\left(-b \frac{\Phi^{3/2}}{E_L}\right)$$

where $\Phi$ is the work function of the emitting material, S is the emitting surface area, $a = 1.54 \times 10^{-6}$ AV$^{-2}$eV and $b = 6.83 \times 10^7$ Vcm$^{-1}$eV$^{-3/2}$ are constants, and $E_L$ is the local electric field that can be expressed as $E_L = \beta V/d$, with $\beta$ the so-called field enhancement factor, i.e. the ratio between the local electric field on the sample surface and the applied field. Accordingly, a linear behaviour is expected for the FN-plot, i.e. $ln(I/V^2)$ $vs$ $1/V$. From the slope $m$ of the FN-plot it is possible to calculate the field enhancement factor as $\beta = -b\,d\,\Phi^{1.5}/m$. The inset of Figure 3(a) shows the FN-plot corresponding to the measured I-V characteristic, and it evidences a clear linearity, confirming that the current is due to the FE phenomenon from the $MoS_2$ surface. Moreover, we can estimate the field enhancement factor as $\beta \approx 40$ (if assuming $\Phi$=5.25eV for the $MoS_2$,[49] and considering the tip correction $k_{tip} \approx 1.5$).

In the same configuration (with $d = 300$ nm) we also tested the FE current stability by applying a constant bias of 45 V and then measuring the emitted current vs time for a period of more than 8 hours. Experimental result is shown in Figure 3(b): a stable current without significant degradation with respect to the average value of about 18 nA is recorded. This observation confirms the suitability of $MoS_2$ flakes for FE applications.

By tuning the cathode-anode separation distance, we characterized the FE properties of the $MoS_2$ flake, in the range 250 nm $< d <$ 2100 nm, in a different location. In Figure 3(c) we show the I-V curves measured for different $d$ values. We observe that, as expected, by increasing the distance of the tip from the surface, higher voltages are necessary to extract electrons from $MoS_2$. The corresponding linear FN-plots, reported in Figure 3(d), confirm the field emission phenomenon. We notice that, for small distances, when high FE current values are obtained (in the range $0.1 - 1$ $\mu A$), further current increase is strongly limited despite the increasing bias voltage, due to a series resistance in the circuit (causing a relevant voltage drop that reduces the local applied field when a high current is flowing) and probably to space charge limited conduction.

A third area of the flake has been also characterized (Figure 3(e)) in a reduced distance range (from 1450 nm to 1750 nm) and with reduced steps in order to precisely analyze the dependence of the turn-on voltage from $d$ in this distance range. As expected, increasing the distance, the emission starts at higher applied voltages, while for reduced distance, it starts at lower voltages. The inset of Figure 3(e) shows the values of the turn-on field(voltage) evaluated for each distance ($E_{ON}, V_{ON}$ $vs$ $d$), evidencing that in such distance range the turn-on voltage is a monotonically raising function of the distance, while the turn-on field is almost saturating at $E_{ON} \approx 26$ V/$\mu$m for $d > 1.5\mu$m.

Finally we report in Figure 3(f) the $\beta$ values extracted from all the reported I-V characteristics discussed above. Interestingly, we observe two different regimes for small and large separation distances. For small distances up to $d \approx 1.5\mu$m, the field enhancement factor increases with the distance, consistently with what has been already reported for FE from $MoS_2$ for very small distances up to 200 nm.[15] Vice versa, for cathode-anode separation greater than 1.5 $\mu$m, we clearly observe that field enhancement factor is rapidly decreasing for increasing distance. We observe that in order to calculate the absolute value of the field enhancement factor it is necessary to infer the effective local work-function of the $MoS_2$ emitting area. However, it has been reported that the work function of layered materials is strongly dependent on the number of layers.[50] Moreover, substrate effects (such as trapped charges) and contamination arising from device processing can significantly modify the work function of $MoS_2$.[51,52] A combined study by using functional scanning probe microscopy techniques and Raman spectroscopy mapping on single and few-layers $MoS_2$ has demonstrated that the work function can vary in the range 4.39 to 4.47 eV depending on the number of layers.[53] Consequently, we report in Figure 3(f) the plot of the extracted $\beta$ values assuming two possible limiting values, $\Phi_{min} = 4.3$ eV and $\Phi_{max} = 5.25$ eV, for the $MoS_2$ work-function in order to take into account possible local variation of this physical property. This simply gives an evaluation of the possible variation of $\beta$ according to the real $\Phi$ value, the overall behavior remaining the same. This variation is observed to be strongly reduced for increasing separation distance.

Our results confirm over a wider range (up to about 1.5$\mu$m) that for small cathode-anode separation distance increasing distance causes an increase of the field enhancement factor, accordingly to previously reported data on FE from $MoS_2$ flakes for very small separation (50 nm < $d$ < 200 nm).[15] Similar behavior has been also reported for several other nanostructures probed at small cathode-anode separation.[54,55] On the other hand, several theoretical and experimental study have confirmed that for larger separation distance (above 1-2 $\mu$m), an opposite behavior is expected, with the field enhancement factor decreasing for raising distance.[56-59]

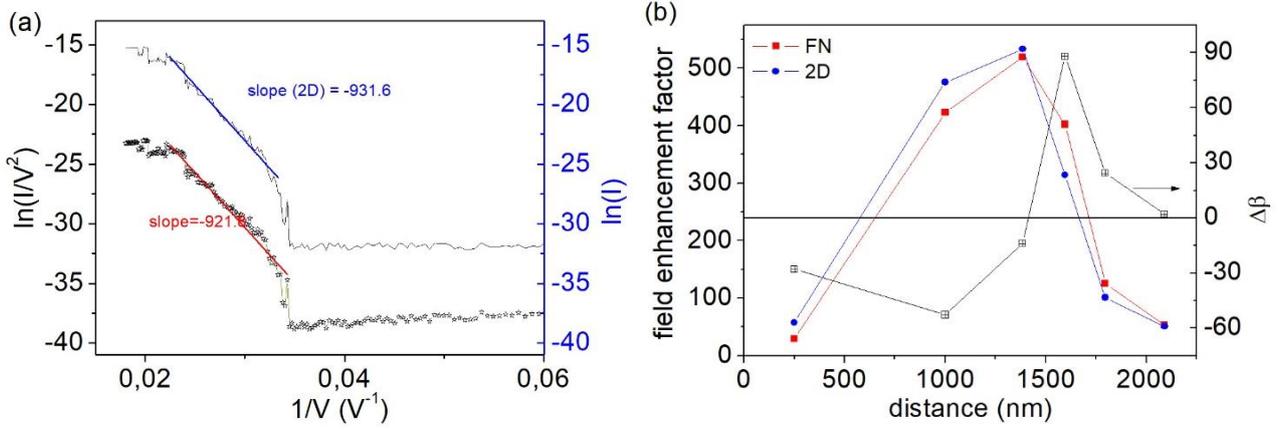

**Figure 4.** (a) Comparison of FN-plot and 2D-FN plot for the FE curve reported in Figure 3(a). Solid lines are the linear fittings for the two plots with the indication of the resulting slopes. (b) Field enhancement factor (left scale) as extracted by the two models, i.e. the standard FN-theory (FN) and the modified model for two-dimensional materials (2D). On the right scale is reported the difference between the values for each distance.

To complete the analysis of the FE properties of few-layer MoS$_2$, we finally analyse the experimental data in the framework of a modified Fowler-Nordheim model for the field-induced vertical electron emission from the surface of 2D materials proposed by Y. S. Ang *et al.* that explicitly takes into account the reduced dimensionality as well as several other effects (non-parabolic energy dispersion, non-conservation of the lateral momentum, finite-temperature and space-charge-limited effects).[60] According to this model (2D-FN), differently by the usual FN-theory, the FE current is described by the formula:

$$I = C \cdot exp\left(-b\frac{\Phi^{3/2}}{E_L}\right)$$

where $C$ is a constant. In this case, it is the plot $ln(I)$ versus $1/V$ that is expected to be linear. In **Figure 4**(a) we show the comparison of the FN-plot and the 2D-FN plot for the FE curve previously shown in Figure 3(a). We notice that from both models a linear behaviour is obtained, and also a similar slope of the linear fitting is extracted. As a consequence, we could not select any of the two models. In Figure 4(b), we report the comparison of the $\beta$ values extracted from the two models (assuming the case $\Phi = 5.25$ eV). The difference between the two values $\Delta\beta = \beta_{FN} - \beta_{FN}^{2D}$ is also reported as a function of the separation distance $d$. Interestingly, for small distances ($d$ <1.5$\mu$m) we found $\beta_{FN} < \beta_{FN}^{2D}$, while for large distances ($d$ >1.5$\mu$m) it is $\beta_{FN} > \beta_{FN}^{2D}$.

## 4. Conclusions

We have used CVD-grown few-layer MoS$_2$ flakes to realize Au/Ti contacted field effect transistor. The characterization of transport properties after electron-beam irradiation for doses up to 100 $e^-$/nm$^2$ has demonstrated an increase of the current in the MoS$_2$ channel as well as a negative shift of the threshold voltage, due to the accumulation of positive charges produced by the irradiation in in SiO$_2$ gate dielectric. We also performed a complete field emission characterization of the same MoS$_2$ flake showing that relative low turn-on field (~20V/$\mu$m) are achievable on few-layer MoS$_2$, making the system of great interest for FE applications, also due to the high current stability demonstrated in our experiment for a period longer than 8 hours.

**References**

[1]   C. Lee, H. Yan, L. E. Brus, T. F. Heinz, J. Hone, S. Ryu, *ACS Nano* **2010**, *4*, 2695.

[2]   B. Radisavljevic, A. Radenovic, J. Brivio, V. Giacometti, A. Kis, *Nature Nanotechnology* **2011**, *6*, 147.

[3]   S. B. Desai, S. R. Madhvapathy, A. B. Sachid, J. P. Llinas, Q. Wang, G. H. Ahn, G. Pitner, M. J. Kim, J. Bokor, C. Hu, H.-S. P. Wong, A. Javey, *Science* **2016**, *354*, 99.

[4]   M.-W. Lin, L. Liu, Q. Lan, X. Tan, K. S. Dhindsa, P. Zeng, V. M. Naik, M. M.-C. Cheng, Z. Zhou, *Journal of Physics D: Applied Physics* **2012**, *45*, 345102.

[5]   X. Tong, E. Ashalley, F. Lin, H. Li, Z. M. Wang, *Nano-Micro Letters* **2015**, *7*, 203.

[6]     A. Di Bartolomeo, L. Genovese, F. Giubileo, L. Iemmo, G. Luongo, T. Foller, M. Schleberger, *2D Materials* **2017**, *5*, 015014.

[7]     B. W. H. Baugher, H. O. H. Churchill, Y. Yang, P. Jarillo-Herrero, *Nano Letters* **2013**, *13*, 4212.

[8]     A. Di Bartolomeo, L. Genovese, T. Foller, F. Giubileo, G. Luongo, L. Croin, S.-J. Liang, L. K. Ang, M. Schleberger, *Nanotechnology* **2017**, *28*, 214002.

[9]     D. Sarkar, W. Liu, X. Xie, A. C. Anselmo, S. Mitragotri, K. Banerjee, *ACS Nano* **2014**, *8*, 3992.

[10]    F. K. Perkins, A. L. Friedman, E. Cobas, P. M. Campbell, G. G. Jernigan, B. T. Jonker, *Nano Letters* **2013**, *13*, 668.

[11]    R. Samnakay, C. Jiang, S. L. Rumyantsev, M. S. Shur, A. A. Balandin, *Applied Physics Letters* **2015**, *106*, 023115.

[12]    L. Yan, H. Shi, X. Sui, Z. Deng, L. Gao, *RSC Advances* **2017**, *7*, 23573.

[13]    H. Yuan, M. S. Bahramy, K. Morimoto, S. Wu, K. Nomura, B.-J. Yang, H. Shimotani, R. Suzuki, M. Toh, C. Kloc, X. Xu, R. Arita, N. Nagaosa, Y. Iwasa, *Nature Physics* **2013**, *9*, 563.

[14]    A. P. S. Gaur, S. Sahoo, F. Mendoza, A. M. Rivera, M. Kumar, S. P. Dash, G. Morell, R. S. Katiyar, *Applied Physics Letters* **2016**, *108*, 043103.

[15]    F. Urban, M. Passacantando, F. Giubileo, L. Iemmo, A. Di Bartolomeo, *Nanomaterials* **2018**, *8*, 151.

[16]    L. Sun, Y. Zhang, G. Hwang, J. Jiang, D. Kim, Y. A. Eshete, R. Zhao, H. Yang, *Nano Letters* **2018**, *18*, 3229.

[17]    W. M. Parkin, A. Balan, L. Liang, P. M. Das, M. Lamparski, C. H. Naylor, J. A. Rodríguez-Manzo, A. T. C. Johnson, V. Meunier, M. Drndić, *ACS Nano* **2016**, *10*, 4134.

[18]    M. Kalbac, O. Lehtinen, A. V. Krasheninnikov, J. Keinonen, *Advanced Materials* **2013**, *25*, 1004.

[19] G. Compagnini, F. Giannazzo, S. Sonde, V. Raineri, E. Rimini, *Carbon* **2009**, *47*, 3201.

[20] F. Giubileo, A. Di Bartolomeo, N. Martucciello, F. Romeo, L. Iemmo, P. Romano, M. Passacantando, *Nanomaterials* **2016**, *6*, 206.

[21] D. Teweldebrhan, A. A. Balandin, *Applied Physics Letters* **2009**, *94*, 013101.

[22] S. Bertolazzi, S. Bonacchi, G. Nan, A. Pershin, D. Beljonne, P. Samorì, *Advanced Materials* **2017**, *29*, 1606760.

[23] B. Peng, G. Yu, Y. Zhao, Q. Xu, G. Xing, X. Liu, D. Fu, B. Liu, J. R. S. Tan, W. Tang, H. Lu, J. Xie, L. Deng, T. C. Sum, K. P. Loh, *ACS Nano* **2016**, *10*, 6383.

[24] H.-P. Komsa, J. Kotakoski, S. Kurasch, O. Lehtinen, U. Kaiser, A. V. Krasheninnikov, *Physical Review Letters* **2012**, *109*.

[25] B. Y. Choi, K. Cho, J. Pak, T.-Y. Kim, J.-K. Kim, J. Shin, J. Seo, S. Chung, T. Lee, *Journal of the Korean Physical Society* **2018**, *72*, 1203.

[26] W. Zhou, X. Zou, S. Najmaei, Z. Liu, Y. Shi, J. Kong, J. Lou, P. M. Ajayan, B. I. Yakobson, J.-C. Idrobo, *Nano Letters* **2013**, *13*, 2615.

[27] C. Durand, X. Zhang, J. Fowlkes, S. Najmaei, J. Lou, A.-P. Li, *Journal of Vacuum Science & Technology B, Nanotechnology and Microelectronics: Materials, Processing, Measurement, and Phenomena* **2015**, *33*, 02B110.

[28] M.-Y. Lu, S.-C. Wu, H.-C. Wang, M.-P. Lu, *Physical Chemistry Chemical Physics* **2018**, *20*, 9038.

[29] O. Ochedowski, K. Marinov, G. Wilbs, G. Keller, N. Scheuschner, D. Severin, M. Bender, J. Maultzsch, F. J. Tegude, M. Schleberger, *Journal of Applied Physics* **2013**, *113*, 214306.

[30] T.-Y. Kim, K. Cho, W. Park, J. Park, Y. Song, S. Hong, W.-K. Hong, T. Lee, *ACS Nano* **2014**, *8*, 2774.

[31] H. Li, Q. Zhang, C. C. R. Yap, B. K. Tay, T. H. T. Edwin, A. Olivier, D. Baillargeat, *Advanced Functional Materials* **2012**, *22*, 1385.

[32] S. Ghatak, A. Ghosh, *Applied Physics Letters* **2013**, *103*, 122103.

[33] A. Di Bartolomeo, A. Grillo, F. Urban, L. Iemmo, F. Giubileo, G. Luongo, G. Amato, L. Croin, L. Sun, S.-J. Liang, L. K. Ang, *Advanced Functional Materials* **2018**, *28*, 1800657.

[34] L. Qi, Y. Wang, L. Shen, Y. Wu, *Applied Physics Letters* **2016**, *108*, 063103.

[35] K. Cho, T.-Y. Kim, W. Park, J. Park, D. Kim, J. Jang, H. Jeong, S. Hong, T. Lee, *Nanotechnology* **2014**, *25*, 155201.

[36] M. Amani, M. L. Chin, A. G. Birdwell, T. P. O'Regan, S. Najmaei, Z. Liu, P. M. Ajayan, J. Lou, M. Dubey, *Applied Physics Letters* **2013**, *102*, 193107.

[37] M.-W. Lin, I. I. Kravchenko, J. Fowlkes, X. Li, A. A. Puretzky, C. M. Rouleau, D. B. Geohegan, K. Xiao, *Nanotechnology* **2016**, *27*, 165203.

[38] F. Giubileo, A. Di Bartolomeo, *Progress in Surface Science* **2017**, *92*, 143.

[39] K. Choi, S. R. A. Raza, H. S. Lee, P. J. Jeon, A. Pezeshki, S.-W. Min, J. S. Kim, W. Yoon, S.-Y. Ju, K. Lee, S. Im, *Nanoscale* **2015**, *7*, 5617.

[40] D. Karmakar, R. Halder, N. Padma, G. Abraham, K. Vaibhav, M. Ghosh, M. Kaur, D. Bhattacharya, T. V. Chandrasekhar Rao, *Journal of Applied Physics* **2015**, *117*, 135701.

[41] J. R. Schwank, M. R. Shaneyfelt, D. M. Fleetwood, J. A. Felix, P. E. Dodd, P. Paillet, Vé. Ferlet-Cavrois, *IEEE Transactions on Nuclear Science* **2008**, *55*, 1833.

[42] A. Di Bartolomeo, A. Scarfato, F. Giubileo, F. Bobba, M. Biasiucci, A. M. Cucolo, S. Santucci, M. Passacantando, *Carbon* **2007**, *45*, 2957.

[43] F. Giubileo, A. D. Bartolomeo, A. Scarfato, L. Iemmo, F. Bobba, M. Passacantando, S. Santucci, A. M. Cucolo, *Carbon* **2009**, *47*, 1074.

[44] F. Giubileo, A. Di Bartolomeo, M. Sarno, C. Altavilla, S. Santandrea, P. Ciambelli, A. M. Cucolo, *Carbon* **2012**, *50*, 163.

[45] F. Giubileo, L. Iemmo, G. Luongo, N. Martucciello, M. Raimondo, L. Guadagno, M. Passacantando, K. Lafdi, A. Di Bartolomeo, *Journal of Materials Science* **2017**, *52*, 6459.

[46] A. Di Bartolomeo, M. Passacantando, G. Niu, V. Schlykow, G. Lupina, F. Giubileo, T. Schroeder, *Nanotechnology* **2016**, *27*, 485707.

[47] L. Iemmo, A. Di Bartolomeo, F. Giubileo, G. Luongo, M. Passacantando, G. Niu, F. Hatami, O. Skibitzki, T. Schroeder, *Nanotechnology* **2017**, *28*, 495705.

[48] R. H. Fowler, L. Nordheim, *Proceedings of the Royal Society A: Mathematical, Physical and Engineering Sciences* **1928**, *119*, 173.

[49] S. Choi, Z. Shaolin, W. Yang, *Journal of the Korean Physical Society* **2014**, *64*, 1550.

[50] S. S. Datta, D. R. Strachan, E. J. Mele, A. T. C. Johnson, *Nano Letters* **2009**, *9*, 7.

[51] O. Ochedowski, K. Marinov, N. Scheuschner, A. Poloczek, B. K. Bussmann, J. Maultzsch, M. Schleberger, *Beilstein Journal of Nanotechnology* **2014**, *5*, 291.

[52] N. D. Kay, B. J. Robinson, V. I. Fal'ko, K. S. Novoselov, O. V. Kolosov, *Nano Letters* **2014**, *14*, 3400.

[53] B. J. Robinson, C. E. Giusca, Y. T. Gonzalez, N. D. Kay, O. Kazakova, O. V. Kolosov, *2D Materials* **2015**, *2*, 015005.

[54] F. Giubileo, A. Di Bartolomeo, L. Iemmo, G. Luongo, M. Passacantando, E. Koivusalo, T. Hakkarainen, M. Guina, *Nanomaterials* **2017**, *7*, 275.

[55] M. Passacantando, F. Bussolotti, S. Santucci, A. Di Bartolomeo, F. Giubileo, L. Iemmo, A. M. Cucolo, *Nanotechnology* **2008**, *19*, 395701.

[56] R. C. Smith, S. R. P. Silva, *Journal of Applied Physics* **2009**, *106*, 014314.

[57] F. Giubileo, A. Di Bartolomeo, L. Iemmo, G. Luongo, F. Urban, *Applied Sciences* **2018**, *8*, 526.

[58] A. Di Bartolomeo, F. Giubileo, L. Iemmo, F. Romeo, S. Russo, S. Unal, M. Passacantando, V. Grossi, A. M. Cucolo, *Applied Physics Letters* **2016**, *109*, 023510.

[59] X. Q. Wang, M. Wang, H. L. Ge, Q. Chen, Y. B. Xu, *Physica E: Low-dimensional Systems and Nanostructures* **2005**, *30*, 101.

[60] L. K. Ang, *arXiv:1711.05898v1*.